\documentclass{iopart}

\usepackage{graphicx}
\usepackage{amssymb}
\usepackage{psfrag}

\begin{document}

\newcommand \be {\begin{equation}}
\newcommand \ee {\end{equation}}
\newcommand \bea {\begin{eqnarray}}
\newcommand \eea {\end{eqnarray}}

\title[Entropic aging and extreme value statistics]{Entropic aging and extreme value statistics}

\author{Eric Bertin}

\address{Universit\'e de Lyon, Laboratoire de Physique, ENS Lyon, CNRS,
46 All\'ee d'Italie, F-69007 Lyon}

\begin{abstract}
Entropic aging consists in a progressive slowing down of the low-temperature
dynamics of a glassy system due to the rarefaction of downwards directions
on the energy landscape, as lower and lower energy levels are reached.
A prototypical model exhibiting this scenario is the Barrat-M\'ezard model.
We argue that in the zero-temperature limit,
this model precisely corresponds to a
dynamical realization of extreme value statistics, providing an interesting
connection between the two fields. 
This mapping directly yields the long-time asymptotic shape of the
dynamical energy distribution, which is then one of the standard
extreme value distributions (Gumbel, Weibull or Fr\'echet), thus restricting
the class of asymptotic energy distributions with respect to
the original preasymptotic results.
We also briefly discuss similarities and differences between the
Barrat-M\'ezard model and undriven dissipative systems like
granular gases. 
\end{abstract}



\section{Introduction}

Glassy systems exhibit a slow relaxation towards their equilibrium state,
giving rise to non-stationary and aging properties \cite{Bouchaud-rev}.
Different physical mechanisms have been put forward to account for
this intriguing phenomenon.
One such mechanism is related to the presence of energy barriers
that the system needs to overcome through thermal activation in
order to explore its phase space. The system is then trapped for long
times in energy wells, and successively explores deeper and deeper traps
\cite{Bouchaud92,Monthus}. In such
a scenario, thermal activation plays a crucial role and the dynamics
becomes frozen in the low temperature limit.
Another standard, though also schematic, mechanism leading to the aging
phenomenon in glassy systems is the so-called 'entropic aging', where
the dynamics is slowed down by the rarefaction of downwards directions.
Energy barriers are then replaced by entropic barriers, and the
temperature no longer plays a prominent role in this case.
Explicit realizations of this scenario have been proposed in simplified
models, like the Backgammon model \cite{Ritort,Franz}, or the
Barrat-M\'ezard model \cite{BarratMezard,Bertin03}, the latter probably being
the simplest model one could think of to illustrate the 'entropic aging'
phenomenon.

In this short note, we establish a direct and simple link between the
Barrat-M\'ezard model and extreme value statistics, namely the statistics
of the maximum or minimum value in a set of random variables
\cite{Gumbel,Galambos}.
This link then allows for the direct application of the standard
results of extreme value statistics to the Barrat-M\'ezard model,
providing in a straightforward way some non-trivial information on
the asymptotic energy distribution reached dynamically in the long-time
limit. We also discuss, in a more exploratory way, a possible qualitative
analogy between the Barrat-M\'ezard model and the free relaxation
of dissipative systems like granular gases.

\section{The Barrat-M\'ezard model}
\label{sec-BM}

As mentioned in the introduction,
the Barrat-M\'ezard model \cite{BarratMezard} is a simple stochastic model
exhibiting the aging phenomenon, due to the presence of entropic barriers.
In a continuous energy formulation, valid in the limit of an infinite
number of microstates, the model is defined as a continuous-time
Markov process in which the system jumps from an energy $E$ to
another energy $E'$ with a Glauber rate:
\be \label{glauber}
w(E'|E) = \frac{\Gamma_0 \,\rho(E')}{1+e^{(E'-E)/T}},
\ee
where $\Gamma_0$ is a microscopic frequency scale,
$\rho(E)$ is the density of microstates at energy $E$,
and $T$ is the heat bath temperature.
The probability $P(E,t)$ to be in a microstate of energy $E$ at time
$t$ obeys the following master equation:
\be
\frac{\partial P}{\partial t}(E,t) = \int_{-\infty}^{\infty}
dE'\, [w(E|E') P(E',t)-w(E'|E) P(E,t)].
\ee
The transition rate (\ref{glauber}) satisfies detailed balance,
so that $P(E,t)$ is expected to converge when $t \to \infty$
to the equilibrium distribution
\be
P_\mathrm{eq}(E,t)= \frac{1}{Z} \, \rho(E)\,e^{-E/T}.
\ee
Depending on the shape of the density of states $\rho(E)$,
the equilibrium distribution $P_\mathrm{eq}(E,t)$ may not be normalizable
below a given temperature \cite{Monthus,Bertin03},
meaning that the continuous energy description breaks down
(the probability measure concentrates on the few lowest energy states).
For instance, if $\rho(E)$ is exponential,
\be \label{dist-exp}
\rho(E)=T_g^{-1}e^{E/T_g} \Theta(-E)
\ee
(with $\Theta$ the Heaviside function), the distribution $P_\mathrm{eq}(E,t)$
is non-normalizable for all temperature $T \le T_g$.
The slow relaxation (or aging) regime can however be described
through a continuous energy description \cite{Bertin03}.
If $\rho(E)$ decays faster than exponentially (e.g., if $\rho(E)$ has
a Gaussian tail), the equilibrium distribution is well-defined
in a continuous energy framework for $T>0$, and aging is interrupted
beyond the equilibration time.

An interesting property of this model is that, contrary to what
happens for trap models involving thermal activation \cite{Bouchaud92,Monthus},
the dynamics is not completely frozen in the zero temperature limit.
The average trapping time $\tau$
needed to escape a given microstate at energy $E$ is given for $T=0$ by
\be
\frac{1}{\tau} = \int_{-\infty}^{\infty} dE'\, w(E'|E)
= \int_{-\infty}^E dE'\,\rho(E').
\label{def-tau}
\ee
Microscopic configurations can then be labelled by $\tau$ rather than
by $E$, and the dynamics of the system is described by the
probability $p(\tau,t)$ to occupy a configuration characterized
by $\tau$ at time $t$.
It has been shown \cite{BarratMezard} that at large time, the dynamical
distribution $p(\tau,t)$ takes a scaling form
\be \label{phi-scal}
p(\tau,t) = \frac{1}{t}\, \phi\left(\frac{\tau}{t}\right),
\ee
with a scaling function $\phi(u)$ given by
\be \label{eq-phi}
\phi(u) = \frac{1}{u^2}\, e^{-1/u}.
\ee
This shape of the scaling function \eref{eq-phi} is generic and
does not depend on the shape of the density of states $\rho(E)$.
However, the dependence on the functional form of $\rho(E)$ is hidden
in the definition of $\tau$ given in Eq.~\eref{def-tau}.
Indeed, expressing $P(E,t)$ as
\be \label{PEt-preas}
P(E,t) = p(\tau,t)\,\left|\frac{d\tau}{dE}\right|,
\ee
one clearly sees from Eqs.~(\ref{def-tau}) and (\ref{phi-scal})
that $P(E,t)$ depends on the detailed shape of $\rho(E)$, suggesting that
Eq.~\eref{PEt-preas} only corresponds to a preasymptotic regime.

Let us also mention that the scaling form (\ref{phi-scal})
is not necessarily restricted to the zero-temperature limit.
In the specific case of an exponential distribution $\rho(E)$,
as given in Eq.~(\ref{dist-exp}),
a similar scaling form exists in the whole aging regime $T<T_g$,
with a temperature-dependent scaling function $\phi$ \cite{Bertin03}.

\section{Mapping onto extreme value statistics}
\label{sec-EVS}

It is interesting to observe that the above zero-temperature dynamics 
can be interpreted as a dynamical realization of extreme value statistics.
To make the analogy more precise, let us first define the discrete time
version of the Barrat-M\'ezard model. The dynamics is defined in the same
way as above, apart from time discretization: at each time step, a new
microstate of energy $E'$ is picked up from the distribution $\rho(E)$,
and the move from $E$ to $E'$ is accepted with probability
\be \label{glauber-dtime}
W(E'|E) = \frac{1}{1+e^{(E'-E)/T}},
\ee
and rejected otherwise.
Consistency between the discrete and continuous time versions of the model
implies that the time step $\tau_0$ in the discretized version should be
chosen as $\tau_0=\Gamma_0^{-1}$, where $\Gamma_0$ is defined in
Eq.~(\ref{glauber}).

It is then easy to see that the zero-temperature discrete-time Barrat-M\'ezard
model exactly implements, in a dynamical way, an extremal process.
Let us assume that at time $t=0$, the initial energy $E_0$ is drawn at random
from the distribution $\rho(E)$, which can be interpreted as
an infinite temperature initial condition (namely, all microstates have
the same weight).
Then at time $t=n\tau_0$, the occupied value $E(t)$ is the lowest value
among the set of $n+1$ random variables $E_j$, $j=0,\ldots,n$,
that have been drawn, and either accepted or rejected, at each time step:
\be
E(t) = \min(E_0,E_1,\ldots,E_n).
\ee
Such a mapping from the Barrat-M\'ezard model to extreme value statistics
is interesting for several reasons. First, it provides an example of a
non-trivial dynamical process which maps exactly onto an extreme value
problem. Second, standard results from the field of extreme value
statistics can then straightforwardly be applied to the Barrat-M\'ezard model,
giving a direct access to the long-time asymptotics of the distribution
$P(E,t)$. 
Indeed, it is known from extreme value statistics
that the shape of $P(E,t)$ asymptotically attains a universal form
belonging to three possible classes, depending on the tail
of the distribution $\rho(E)$ \cite{Gumbel,Galambos}. 
These universal distributions are obtained through an asymptotic rescaling,
\be \label{eq-rescal}
P(E,t) = \frac{1}{b_t}\; \Phi\left(\frac{E-a_t}{b_t}\right),
\quad t \gg \tau_0,
\ee
with suitably chosen time-dependent parameters $a_t$ and $b_t$.
If $\rho(E)$ decays for $E \to -\infty$
faster than any power law, then the asymptotic distribution $\Phi(z)$ has
a Gumbel (or Fisher-Tippett-Gumbel) form, namely
\be \label{dist-gumbel}
\Phi(z) = \Phi_\mathrm{g}(z) \equiv e^{z-e^z}.
\ee
Note that we consider here the statistics of minimum values, so that the
definition (\ref{dist-gumbel}) of the Gumbel distribution differs by a
sign reversal from the more usual definition related to maxima.

If $\rho(E)$ instead decays as a power law $\rho(E)\sim 1/|E|^{1+\alpha}$
($\alpha>0$) when $E \to -\infty$, then $\Phi(z)$ is the Fr\'echet distribution
\be
\Phi(z) = \Phi_\mathrm{f}(z) \equiv \frac{\alpha}{|z|^{1+\alpha}}\, e^{-|z|^{-\alpha}}
\qquad (z<0).
\ee
Finally, in the case when $E$ is bounded from below, and $\rho(E)$
behaves as a power law close to the bound $E_\mathrm{min}$,
\be \label{weibull-class}
\rho(E) \sim (E-E_\mathrm{min})^{\nu-1}, \qquad E \to E_\mathrm{min}
\quad (E>E_\mathrm{min})
\ee
with $\nu>0$, the asymptotic distribution is the Weibull one
\be
\Phi(z) = \Phi_\mathrm{w}(z) \equiv \nu z^{\nu-1}\, e^{-z^{\nu}} \qquad (z>0).
\ee
These three cases can be encompassed into a single distribution
$\Phi_{\gamma}(z)$, using a slightly different choice of the rescaling
parameters $a_t$ and $b_t$ \cite{vonMises}
\be \label{eq-mises}
\Phi_{\gamma}(z) = (1-\gamma z)^{-1-\frac{1}{\gamma}}\,
\exp\left[-(1-\gamma z)^{-\frac{1}{\gamma}}\right],
\qquad 1-\gamma z >0,
\ee
with a real parameter $\gamma$. By continuity, 
$\Phi_{\gamma}(z)$ should be interpreted for $\gamma=0$
as $\exp[-\exp(z)]$, which corresponds to the Gumbel
distribution. For $\gamma>0$, Eq.~(\ref{eq-mises}) corresponds to the
Fr\'echet case, with $\gamma=1/\alpha$.
For $\gamma<0$, the Weibull distribution
is obtained, with $\gamma=-1/\nu$.

As a result, we have shown using a mapping onto extreme value statistics
that the dynamical energy distribution $P(E,t)$ of the Barrat-M\'ezard model
asymptotically corresponds to one of the standard extreme value distributions
(Gumbel, Weibull or Fr\'echet), depending on the behaviour of the tail of the 
density of states $\rho(E)$.
This result significantly restricts the class of allowed asymptotic
distributions with respect to the original results obtained
by Barrat and M\'ezard. As mentioned at the end of Sec.~\ref{sec-BM},
the distribution $P(E,t)$ deduced from the scaling form \eref{phi-scal}
actually corresponds to a family of distributions indexed by the full
density of states $\rho(E)$, once the variable $\tau$ is expressed as
a function of $E$.
In constrast, the present distributions obtained from extreme value
statistics correspond to a one-parameter family of distributions,
that is to a much more restricted class.
This difference can be ascribed to the fact that the scaling regime
\eref{phi-scal} can be considered as a preasymptotic regime.

The above results are valid, strictly speaking, in the case of
discrete time dynamics. However, the type of dynamics (discrete
or continuous time) should not influence the large-time asymptotic
dynamics of the system, so that extreme value distributions
are expected to hold also for continuous time dynamics.
This is confirmed by the simple example
of the exponential distribution (\ref{dist-exp}),
which falls into the Gumbel class. In this specific case,
one has $\tau(E)=e^{-E/T_g}$, and the scaling form (\ref{phi-scal})
with $\phi$ given by (\ref{eq-phi}) directly
boils down to the Gumbel distribution.

\section{Illustration on large systems}

In this section we briefly illustrate the above results on two different
classes of systems: mean-field models like the Random Energy Model
(REM) \cite{Derrida,BouchMez97}, and finite-dimensional models with
short-range correlations.

In the REM, which can be considered as the simplest disordered system,
all microscopic configurations have an energy $E$ drawn at random from
a Gaussian distribution $\rho(E)$ with a variance proportional to $N$
\cite{Derrida}. Given that the low-temperature properties of the model
depend on the low-lying energy states, only the low energy tail matters.
It has thus been proposed \cite{BouchMez97} to
generalize the Gaussian distribution to a distribution $\rho(E)$
with a tail
\be
\rho(E) \sim e^{-|E|^{\beta}/N^{\beta-1}}, \qquad E \to -\infty
\quad (\beta >1),
\label{rhoE-REM}
\ee
where $N \gg 1$ is the underlying number of degrees of freedom.
Algebraic prefactors can also be included, but they do not influence
the asymptotic behavior \cite{BouchMez97}.
The REM exhibits interesting equilibrium properties, like a freezing
transition at a finite temperature $T_c=\beta^{-1} (\ln 2)^{(1-\beta)/\beta}$
\cite{BouchMez97}.

In its original formulation, the REM is an equilibrium model, without
any specified dynamics. A rather natural choice is to use
a Glauber dynamics, connecting all microscopic configurations
with the probability \eref{glauber-dtime}.
It has been shown that for a finite number of states and a non-zero
temperature, the dynamics
of the REM becomes similar to that of the trap model beyond a
size-dependent time scale, which diverges with system size \cite{BenArous}.
However, in the zero-temperature and infinite size limits,
the dynamics by definition corresponds to the Barrat-M\'ezard model,
with the density of states $\rho(E)$ given in Eq.~\eref{rhoE-REM}.
From the results of Sec.~\ref{sec-EVS}, the corresponding long-time energy
distribution $P(E,t)$ is a Gumbel distribution, for all $\beta >1$,
since the density of states $\rho(E)$ decays faster than any power law.

As a second example, we consider the case of a system composed
of $N$ independent degrees of freedom, in the sense that the energy
can be decomposed into a sum of $N$ independent terms associated
to each degree of freedom. In many cases, the density of states $\rho(E)$
of such systems can be written in the form
\be
\rho(E) \approx C (E-E_\mathrm{min})^{\eta N -1}
\ee
for $E$ close enough to $E_\mathrm{min}$ ($E>E_\mathrm{min}$),
where $C$ and $\eta$ are positive constants.
For instance, in a perfect gas of $N$ monoatomic particles, one has
$E_\mathrm{min}=0$ and $\eta=d/2$, where $d$ is the space dimension.

We now determine the dynamical energy distribution $P(E,t)$,
again from the results of Sec.~\ref{sec-EVS}.
The energy density $\rho(E)$ satisfies Eq.~(\ref{weibull-class})
with $\nu=\eta N$.
Hence the dynamical energy distribution falls within
the Weibull class.
The asymptotic distribution is then given by Eq.~(\ref{eq-mises})
with a parameter $\gamma=-1/\nu=-1/(\eta N)$, which is very small
for large $N$.
We know that the distribution (\ref{eq-mises}) converges to the Gumbel
law when $\gamma \to 0$, thus for large $N$ it becomes
very close to the Gumbel distribution (\ref{dist-gumbel}).

Hence, the connection to extreme value statistics presented in this paper
allows one to determine in a straightforward way the dynamical
energy distribution of the Barrat-M\'ezard model with an arbitrary
density of states. In addition, it turns out that in many situations
of interest, the dynamical energy distribution $P(E,t)$ of a large system
is actually very close to the Gumbel distribution, as can be seen on the
above two examples.

\section{Discussion: from aging to relaxation in dissipative models}

\subsection{A toy model for undriven dissipative systems}

The Barrat-M\'ezard model has been defined within the framework of aging
systems, in the zero-temperature limit.
During the relaxation process, energy is progressively released to the
zero-temperature reservoir. From a theoretical point of view, this situation is
reminiscent of the dynamics of freely cooling dissipative systems,
like undriven granular gases, or freely decaying turbulence.
In such systems, the elementary degrees of freedom retained in the
description are macroscopic entities (grains, etc.) or collective modes
(for instance Fourier modes) coupled via dissipative interactions.
In the example of the granular gas, energy is dissipated through
collisions between grains, while in the case of turbulence, dissipation
results from the viscous damping.

In most realistic cases, dissipative events like collisions involve only a
small number of degrees of freedom in the systems, and thus do not lead
to large scale changes of the configuration of the system.
However, in a mean-field spirit, one could model a dissipative event
as a transition to a randomly chosen configuration with lower energy.
Although very naive, this picture however leads to a reasonable
scaling of the dissipated energy in each event, as a function of system
size. Indeed, one expects in a gas composed of $N$ grains that the fraction
of energy dissipated in a collision is proportional to $1/N$,
since only a finite number of grains (typically two) are involved in
the collision.
Considering now the Barrat-M\'ezard model defined with the energy density
$\rho(E) \propto E^{\frac{dN}{2}-1}$, corresponding to a gas of particles,
we compute the conditional average value of the energy $E'$
reached after a single transition starting from energy $E$, yielding
\be
\langle E' \rangle_E = \frac{\int_0^E dE'\, E' \rho(E')}{\int_0^E dE'\, \rho(E')} = \frac{\frac{dN}{2}}{\frac{dN}{2}+1} \, E.
\ee
It follows that the typical amount $\Delta E_\mathrm{typ}$ of energy
dissipated to the zero-temperature heat bath reads, in the large $N$ limit
\be
\Delta E_\mathrm{typ} \equiv E-\langle E' \rangle_E \approx \frac{2E}{dN}.
\ee
Hence, despite its mean-field nature, the Barrat-M\'ezard dynamics yields
the correct scaling for the energy dissipated in a single event.
This result suggests that some qualitative features of dissipative
systems might be captured by the Barrat-M\'ezard model.

\subsection{Global fluctuations in granular gases}

Along this line of thought, and given the connection presented in
Sec.~\ref{sec-EVS} between extreme value statistics
and the Barrat-M\'ezard model, it is natural to wonder whether
the latter model could qualitatively account for the surprising emergence
of distributions close to the Gumbel form in the relaxation regime
of granular gases \cite{Brey}.
More precisely, it was found in numerical simulations of a two-dimensional
freely cooling granular gas that the shape of the energy distribution
$P(E,t)$ is well-described by a Gumbel-like distribution \cite{Brey},
also called BHP-distribution in this context \cite{Bramwell}.

However, the Gumbel distribution observed in the Barrat-M\'ezard model
differs from that reported in granular gases \cite{Brey} by a sign reversal.
The former, being the distribution of minimum values, has an exponential
tail on the left side, while the latter exhibits an exponential tail
on the right side.
Hence, the simple cooling mechanism at play in the Barrat-M\'ezard model
is not responsible for the appearance of the BHP-distribution
in undriven dissipative systems. The reason for the appearance of
Gumbel-like distributions rather seems to lie in the criticality
of the dynamics due to the proximity of an instability
threshold \cite{Brey,Brey2}, as such distributions
were originally reported in critical systems \cite{Bramwell,Racz1,Racz2}.

\section{Conclusion}

In this short note, we have illustrated how the discrete-time
Barrat-M\'ezard model at zero temperature,
a simple model exhibiting 'entropic' aging,
is quantitatively related to the extreme value statistics of
independent and identically distributed random variables.
Such a connection provides interesting insights into the long time statistics
of the Barrat-M\'ezard model, showing that the dynamical energy
distribution asymptotically corresponds to one of the limit distributions
arising in extreme value statistics (Gumbel, Fr\'echet or Weibull).
This mapping has been established in the framework of the discrete-time
model, but for times $t \gg \tau_0$ the time discretization becomes
irrelevant, and the results for the long-time behaviour of
the distribution $P(E,t)$ should also hold for the continuous-time
model.
Note also that the connection presented here between glassy systems 
and extreme value statistics differs from that reported in
\cite{BouchMez97}.

An interesting issue to be explored further would be the link to
record breaking processes \cite{Krug}, that have been put forward
as an important mechanism at play in aging dynamics \cite{Sibani}.
In the Barrat-M\'ezard model at zero temperature, all transitions
occuring between configurations can be seen as record breakings,
if one considers that random numbers are drawn at each time-step of
the discrete-time dynamics, as proposed in Sec.~\ref{sec-EVS}.
Another line for further investigation deals with possible connections
between extreme value statistics
and problems of sums of random variables. Such a connection has
already been established in a different context \cite{Bertin05,BC06,CB08}.
There, the maximum in a set of independent and identically distributed
random variables has been mapped onto a sum of non-identically distributed,
and most often correlated, random variables. Such a mapping
accounts for instance for the appearance of the Gumbel distribution in the
statistics of the integrated spectrum in $1/f$-noise problems
\cite{Antal01,Bertin05}.
In the Barrat-M\'ezard model, an analogous connection could possibly be
made. Indeed, the energy reached at time $t$ can be expressed as the sum
of the energy at time $t=0$ and of all the energy jumps between times
$0$ and $t$. However, the number $n$ of terms in the sum (that is, the number
of jumps) is not constant, and one expects $n$ to be correlated in a
non-trivial way to the total energy $E$, making the problem difficult
to handle from the probabilistic side.
Nevertheless, results obtained in a similar spirit for extreme values
of sets of a non-constant number of random variables have
been reported recently \cite{Godreche}, suggesting that
such statistical issues should be worth investigating.

\end{document}